\documentstyle[epsfig, twocolumn]{esapub}

\begin{document}

\setlength{\parindent}{0pt}
\setlength{\parskip}{ 10pt plus 1pt minus 1pt}
\setlength{\hoffset}{-1.5truecm}
\setlength{\textwidth}{ 17.1truecm }
\setlength{\columnsep}{1truecm }
\setlength{\columnseprule}{0pt}
\setlength{\headheight}{12pt}
\setlength{\headsep}{20pt}
\pagestyle{esapubheadings}

\title{\bf TOPOLOGICAL CONSTRAINTS ON THE RELAXATION OF COMPLEX 
MAGNETIC FIELDS}

\author{{\bf G.~Hornig } \vspace{2mm} \\
Theoretische Physik IV, Ruhr-Universit\"{a}t Bochum,
              44780 Bochum, Germany \\ (email: gh@tp4.ruhr-uni-bochum.de)}

\maketitle

\begin{abstract}
Newly emerging magnetic flux can show a complicated linked or
interwoven topology of the magnetic field. The complexity of this linkage or
knottedness of magnetic flux is related to the free energy stored in the
magnetic field. Magnetic reconnection provides a process to release
this energy on the time scale of the dynamics. At the same time it
approximately conserves the total magnetic helicity. Therefore the 
conservation of total magnetic helicity is a crucial constraint for
the relaxation of complex magnetic fields. However, the total magnetic
helicity is only the first, most elementary, quantity of an infinite 
series of topological invariants of the magnetic field. All these
invariants are strictly conserved in ideal magnetohydrodynamics. As an 
example a preliminary set of these invariants is derived. The
relevance of these higher order invariants for the final state of
relaxation under magnetic reconnection
and their implications for the release of magnetic energy are discussed. 


  Key~words: magnetic fields; topological invariants; magnetic reconnection.

 \end{abstract}

\section{INTRODUCTION}
\begin{figure}[h]
   \begin{center}
   \leavevmode
   \centerline{\epsfig{file=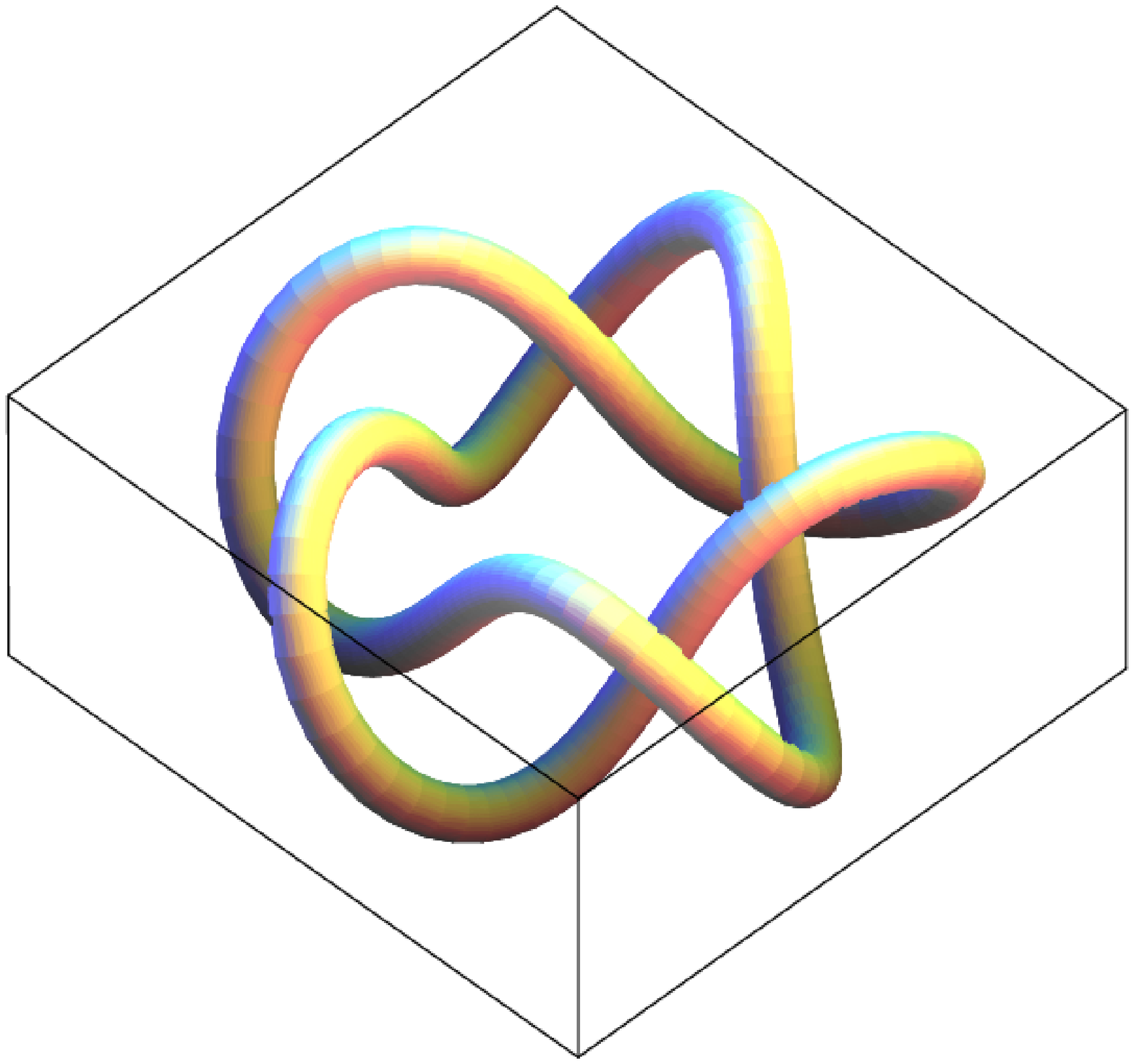,width=3.5cm,height=2.8cm}a) 
                \epsfig{file=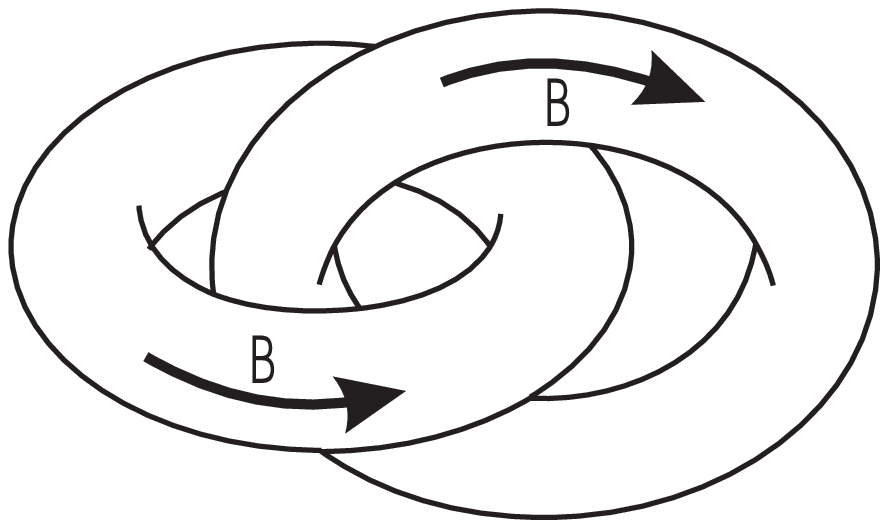,width=3cm,height=2.5cm}b)}
   \end{center}
 \caption{\em (a) A thin flux tube forming a torus knot of type
   (n=2,m=5) and (b) a simple linkage of two magnetic flux tubes, which
   leads to a  non-vanishing total helicity and thus to a lower bound of the
   magnetic energy.}

 \label{fig1}
 \end{figure}

Magnetic fields in the corona often show a non-trivial topology,
that is the magnetic field lines are interwoven or knotted. They form, 
in mathematical terms ``knots'' and ``links''. Here  ``knot''
refers to a single field line of non-trivial topology, while ``link'' 
is used if there are at least two field lines which can not be
separated without cutting of lines. Instead of using isolated field
lines we can apply these notions to flux tubes as well. Note that the existence of complicated knots and linkages of magnetic
flux is in no way artifical. Consider for instance a closed magnetic 
flux tube. In magnetostatic equilibrium the surfaces of constant
pressure are magnetic surfaces, that is the magnetic field is
everywhere tangent to the surface. Depending on the ratio of poloidal
to toroidal component  the field lines will close
after a certain number of windings along the flux tube ($n$) and
around the core of
the flux tube ($m$), or they will  ergodically fill the
whole flux surface. In the former case the flux surface is called a 
rational surface, while in the 
latter case its called an irrational surface according to the terminology in tokamak physics. In generic cases
each of the infinite set of these rational surfaces has different
ratio $n/m$ and therefore for each rational surface the field lines
are knotted in form of a so called torus knot of type (n,m). A simple example of these torus knots is shown in Figure~\ref{fig1}a). 

Magnetic fields which show knotted or linked magnetic flux contain a
certain amount of free energy. Here the notion  ``free energy'' denotes 
the difference
between   the magnetic energy ${\it E}(B) = \int_V
B^2/(8 \pi) d^3r$ of initial configuration and the lowest possible magnetic energy,
that is the energy of a vacuum magnetic field satisfying the same boundary
conditions. This free energy is  stored in the magnetic field 
and could be set free in a relaxation process. The process of
relaxation can be formally separated into two subsequent
processes. First an ideal relaxation, that is a relaxation under an
ideal Ohm's law, which conserves the magnetic topology and leads to a
lower magnetic energy. Second a
non-ideal relaxation, for instance by magnetic 
reconnection, which might finally lead to the vacuum field and thus to 
the lowest energy state. Of course in reality most likely the reconnection
process will set in before the lowest energy state accessible under an 
ideal evolution is reached. Moreover, it will probably  not
directly lead to the vacuum field  but to a higher energy state, which again
might be the starting point of a subsequent relaxation process and
 might involve further reconnection processes. But
nevertheless the difference between the lowest energy state accessible 
by an ideal evolution and the energy of the corresponding vacuum
state is well determined and is refered to as minimum free energy. It
is only determined by the topology of the initial state and 
boundary conditions.

During a non-ideal relaxation the minimum free energy can be converted
in thermal or kinetic energy of the plasma. Thus it might be the source of the energy which is required to heat
the solar corona and the determination of the free energy could  
give an important estimate for the energy available for this
process. In the next section we will show the relation
between the free energy and the magnetic topology which should
motivate the investigation of higher order topological invariants. A
preliminary version of such measures of the magnetic topology  will be given in Section~3, while the third
section is devoted to the evolution of these invariants under reconnection.

\section{LOWER BOUNDS FOR THE FREE ENERGY} 
Roughly speaking the minimum free energy
increases with the complexity of the magnetic field. This qualitative
statement was given a quantitative meaning in several papers, e.g.
 \cite*{Arn}, \cite*{Freed1}, \cite*{Freed2}, \cite*{Ber1}, \cite*{Freed3}, showing
for instance that the total magnetic helicity, 
\begin{equation}
H({\bf B}) = \int_V {\bf A}\cdot {\bf B} \ d^3x \label{hel}
\end{equation}
can provide such a lower bound. (For a detailed discussion of magnetic helicity see \cite{Chap}).
 In the simplest case, that 
is if the magnetic field is enclosed in the volume $V$ (no magnetic
flux is penetrating the boundary $\partial V$ of the volume , ${\bf B} \cdot
{\bf n} |_{\partial V}=0$), this reads 
\begin{equation}
{\it E}(B) \ge C |H(B)|, \label{helbound}
\end{equation}
where $C$ is a constant depending only on the shape and size of the
volume $V$. Hence, if the magnetic field has a non-vanishing total
helicity, resulting for instance from a simple linkage of two flux
tubes as shown in Figure~\ref{fig1}b), the magnetic energy can not decrease 
below this bound as long as the topology of the field is conserved.  A lower
energy can only be obtained in an evolution which changes the magnetic 
topology, like for instance magnetic reconnection. The 
conservation of topology is usually provided by the ideal evolution of 
the plasma (${\bf E} + {\bf v} \times {\bf B} =0$), although the
condition is more general and there are several cases of  non-ideal evolution 
which do not change the magnetic topology (\cite{Horn1}).

\begin{figure}[h]
   \begin{center}
   \leavevmode
   \centerline{{\epsfig{file=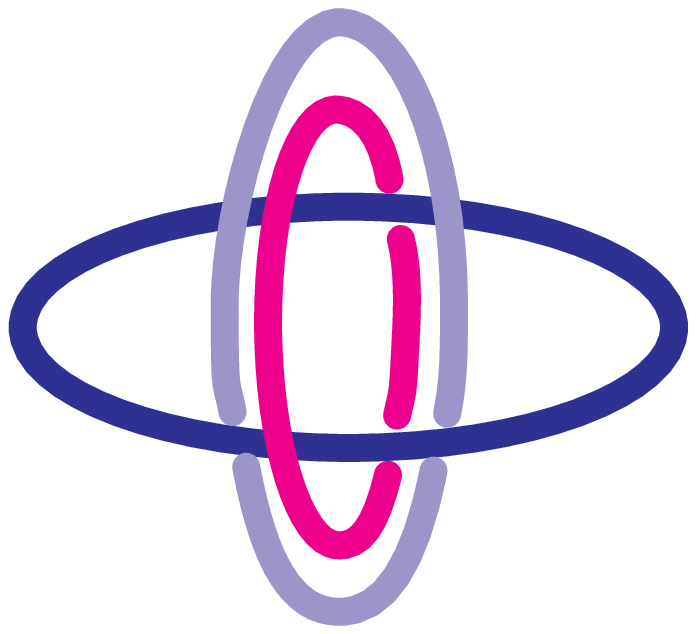,
               width=2.5cm}}a)   \epsfig{file=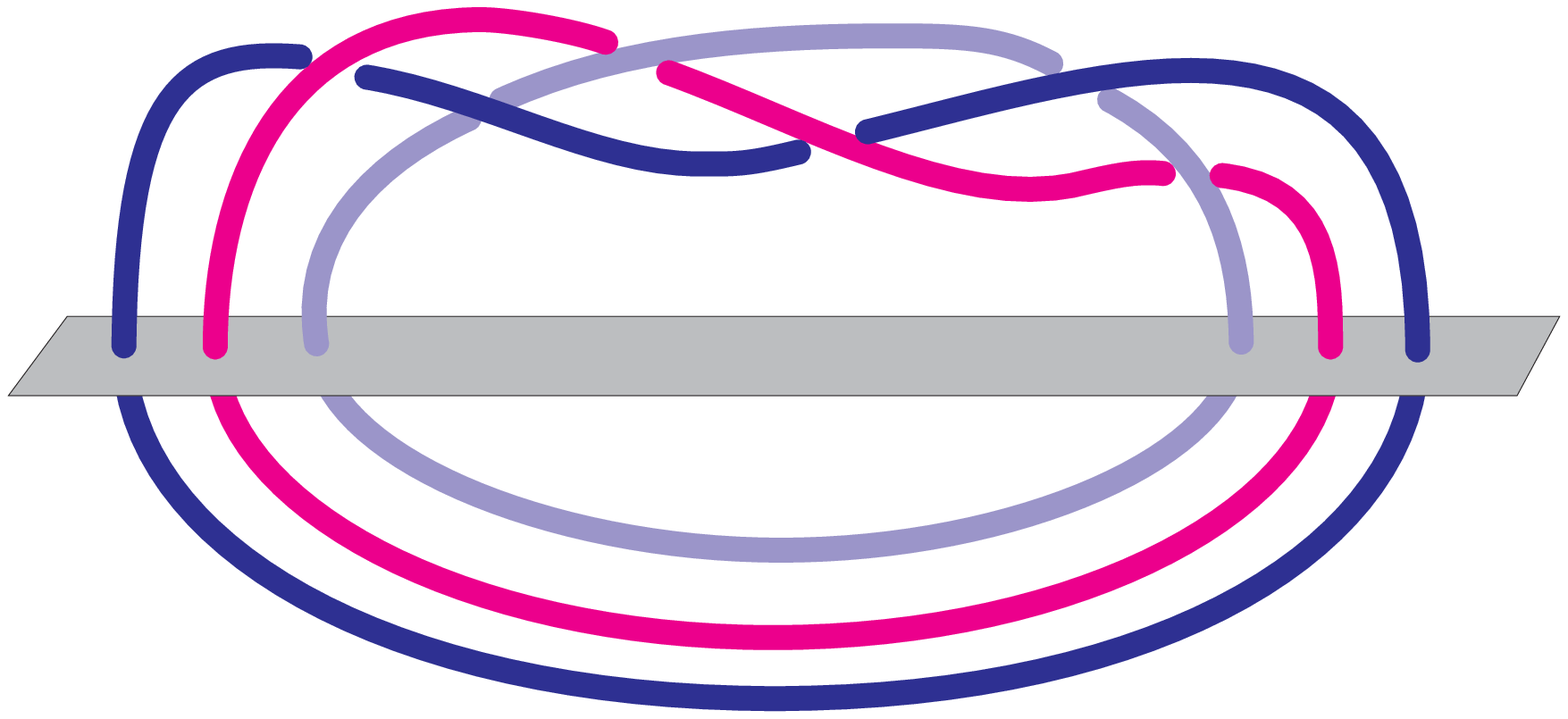,
               width=5.0cm}b)}
   \end{center}
 \caption{\em (a) The Borromean rings. A configuration of vanishing total
   helicity but non-trivially linked. (b) An topological equivalent
   configuration in form of braided flux tubes. }

 \label{fig2}
 \end{figure}

However, the total helicity is only a very rough measure of the
topology of a configuration and the lower bound given in Eq.~\ref{helbound}
might be very low or even zero although the topology of the field is
non-trivial. An example for a configuration which has a vanishing
total magnetic helicity but is non-trivially linked are the so called 
Borromean rings shown in Figure~\ref{fig2}a) and in a topological equivalent
configuration in Figure~\ref{fig2}b). For this configuration the lower 
bound given by (\ref{helbound}) is zero, but a
generalization of such an inequality was given by \cite*{Freed3} with
the help of the so called asymptotic crossing number, which shows that 
there exists a non-zero lower bound for the energy in this case as well. 

A drawback of the use of the so called asymptotic crossing number is
its complicated and abstract definition. It requires the decomposition 
of the field in closed flux tubes (which is not possible in general)
and a complicated minimization process applied to each combination of
linked flux tubes. While this number is important to prove the
existence of lower bounds of the energy, it is impossible to calculate it 
for a generic magnetic field. This is in sharp contrast to the total
magnetic helicity, which can easily be calculated for arbitrary
magnetic fields irrespective of whether a decomposition into 
flux tubes exists or not. 

This is the motivation to look for invariants, analogous to magnetic helicity, which give non-vanishing  values for higher forms
of knottedness or linkage of magnetic flux in ideal (topology
conserving) evolutions. These higher order topological invariants would provide
inequalities similar to Eq.~\ref{helbound}. An additional reason for 
the search for such invariants is that they would lead to a deeper
understanding of the global properties of magnetic fields.

\section{HIGHER ORDER TOPOLOGICAL INVARIANTS}
A formula for higher order topological invariants analogous to the
magnetic helicity does not exist yet. However, there are constructions 
of such invariants if the magnetic field is confined to a set of 
isolated magnetic flux tubes. Such a formula was given for instance
in \cite*{Ber2} for a third order invariant and in \cite*{Ruz} for a
forth order invariant. Note that we are looking for invariants which
depend on the magnetic field only, while in the presence of other conserved
quantities, such as mass density or entropy, it is possible to combine 
these and in this way  construct new invariants (\cite{Tur}).

 In the following a method is provide to
construct an infinite sequence of higher order invariants which can
be calculated for arbitrary magnetic fields, but have the drawback that 
they require a specific gauge. Therefore, they should be considered as
a preliminary form, which, however, allows us to gain insight in the
properties of such invariants. 

The invariants we are looking for are of the form 
\begin{equation}
H^{(n)} = \int_V h^{(n)}({\bf B}) \ d^3r
\end{equation}
for a magnetic field ${\bf B}(x,t)$ enclosed in the (simply connected) volume $V$. These are topological  invariants if for a frozen-in magnetic field, 
\begin{equation}
\partial_t {\bf B} - \nabla \times \left({\bf v} \times  {\bf
    B}\right)  = 0, \label{Bcons}
\end{equation}
the densities $ h^{(n)}$ satisfy 
\begin{equation}
\partial_t h^{(n)} + \nabla\! \cdot \!({\bf v} \  h^{(n)}) = 0.
\label{helcons}
\end{equation}
This results for the integral in
\begin{equation}
\frac{d}{dt} H^{(n)} = 0, \label{Helcons}
\end{equation}
for a comoving volume $V$. It is easy to check that Eq.~\ref{helcons}
holds for $h^{(n)}$ defined by 
\begin{equation}
h^{(n)} := {\bf A}\cdot{\bf G}^{(n)} \label{hdef}
\end{equation}
 provided the vector fields ${\bf A}(x,t)$ and  ${\bf G}(x,t)$ satisfy
\begin{eqnarray}
& \partial_t {\bf A} + \nabla  \left( {\bf v \! \cdot  \! A}\right)
                - {\bf v} \times (\nabla \times {\bf A})  = 0 \label{Aeqn}\\
& \partial_t {\bf G}^{(n)}  - \nabla \times \left({\bf v} \times 
                {\bf G}^{(n)}\right) + {\bf v} \ \nabla\! \cdot \!{\bf
                G}^{(n)} = 0. \label{Geqn}
\end{eqnarray}
Now, if ${\bf A}$ is the vector potential of ${\bf B}$,  Eq.~\ref{Bcons} 
yields 
\begin{equation}
\partial_t {\bf A}  - {\bf v} \times \nabla \times {\bf A}  = \nabla \chi.
\end{equation}
Thus we can meet Eq.~\ref{Aeqn} with the gauge 
${\bf \tilde A}(x,t) = {\bf  A}(x,t) + \nabla \Psi(x,t)$  defined by 
\begin{equation}
 \partial_t \Psi =  -\chi - {\bf v}\! \cdot \!({\bf A}+\nabla \Psi) 
\end{equation}
Note that this equation can be integrated in time for an arbitrary
initial gauge $\Psi(x,0)$.
Similar, defining ${\bf G}^{(n)}$ by 
\begin{equation} 
\nabla \cdot {\bf G}^{(n)} := h^{n-1} \label{Gdef}
\end{equation}
Eq.~\ref{helcons} for $n-1$ results in 
\begin{equation}
 \partial_t {\bf G}^{(n)}  + {\bf v} \ \nabla\! \cdot \!{\bf  G}^{(n)} =
 \nabla \times {\bf F}
\end{equation}
for some vector field ${\bf F}(x,t)$. Again we can meet Eq.~\ref{Geqn}
using a gauge ${\bf \tilde G}^{(n)}(x,t) = {\bf  G}^{(n)}(x,t) + \nabla \times {\bf
  J}(x,t)$ defining ${\bf J}$ by
\begin{equation}
\partial_t {\bf J} :=  - {\bf F} + {\bf v} \times ( {\bf
  G}^{(n)} + \nabla \times {\bf J}), 
\end{equation}
and the gauge at an initial time is free (${\bf J}(x,0)$), but
the equation determines the gauge for all later times.

We can now rename ${\bf \tilde A}$ by ${\bf A}$ and ${\bf \tilde G}$
by ${\bf G}$. Starting  with ${\bf G}^{(2)} = {\bf B}$ equations (\ref{hdef}) and
(\ref{Gdef}) yield  an infinite recursively
defined  sequence of integrals of the magnetic field, all of which satisfy Eq.~\ref{Helcons}. The first ($H^{(2)}$) is the total
magnetic helicity, which is a second order invariant in that it
depends quadratically on ${\bf B}$
\begin{equation}
H^{(2)} = \int_V {\bf A}\cdot{\bf B} \ d^3r.
\end{equation}
The next invariant is of third order
\begin{equation}
H^{(3)} = \int_V {\bf A}\cdot{\bf G} \ d^3r \quad \mbox{with} \quad 
\nabla \cdot {\bf G}^{(n)} := {\bf A}\cdot{\bf B},
\end{equation}
and correspondingly $H^{(n)}$ is of order $n$.
The drawback of this construction is the choice of the special gauge
for ${\bf A}$ and ${\bf G}$, which for instance does not allow to
integrate  ${\bf A}$ by
\begin{equation}
{\bf A}= \int_V \nabla_r \frac{1}{|{\bf r} -{\bf r'}|} \times {\bf
  B}({\bf r'}) \ d^3r',
\end{equation}
since this corresponds to a different gauge ($\nabla \cdot {\bf A}
=0$), which in general does not satisfy Eq.~\ref{Aeqn}. Thus, the value of
$H^{(n)}$ is not uniquely determined, but constant for a frozen-in
magnetic field.

\section{RELAXATION UNDER RECONNECTION}
\begin{figure}[h]
   \begin{center}
   \leavevmode
   \centerline{\epsfig{file=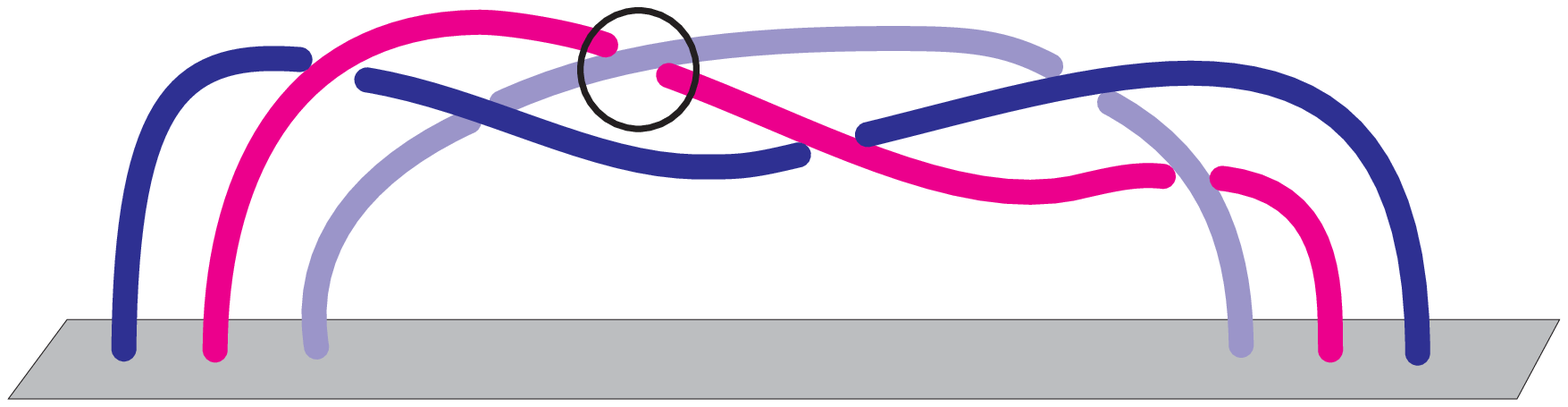,
               width=6.0cm}}
\centerline{\epsfig{file=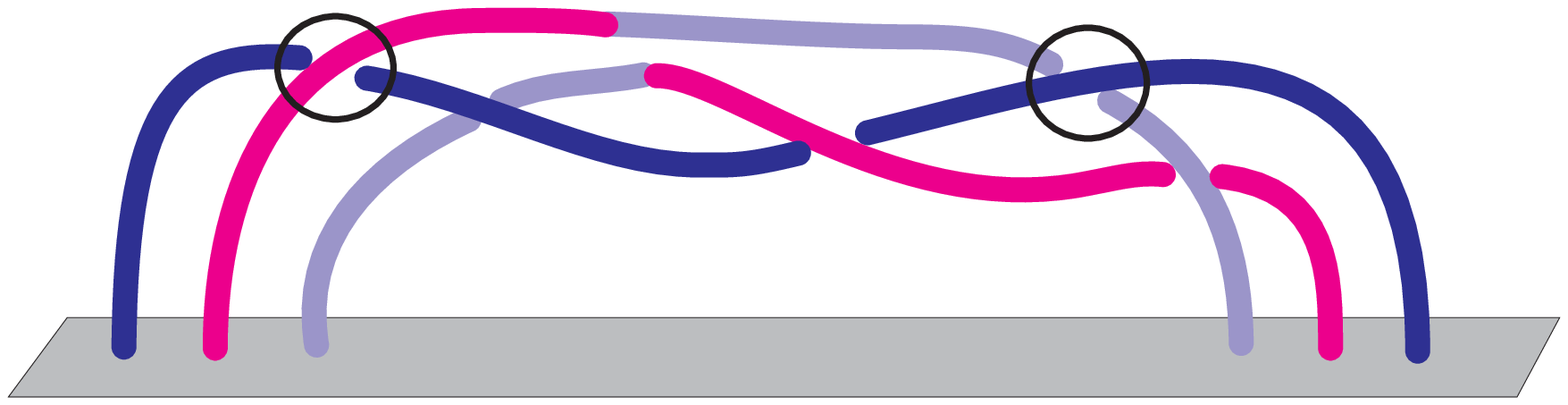,
               width=6.0cm}}
\centerline{\epsfig{file=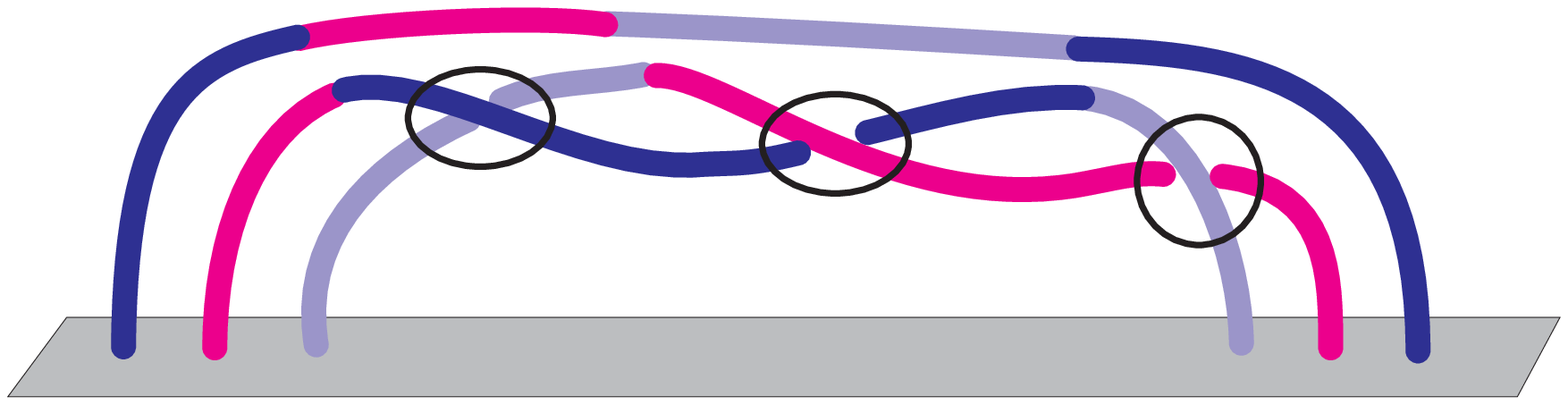,
               width=6.0cm}}
\centerline{\epsfig{file=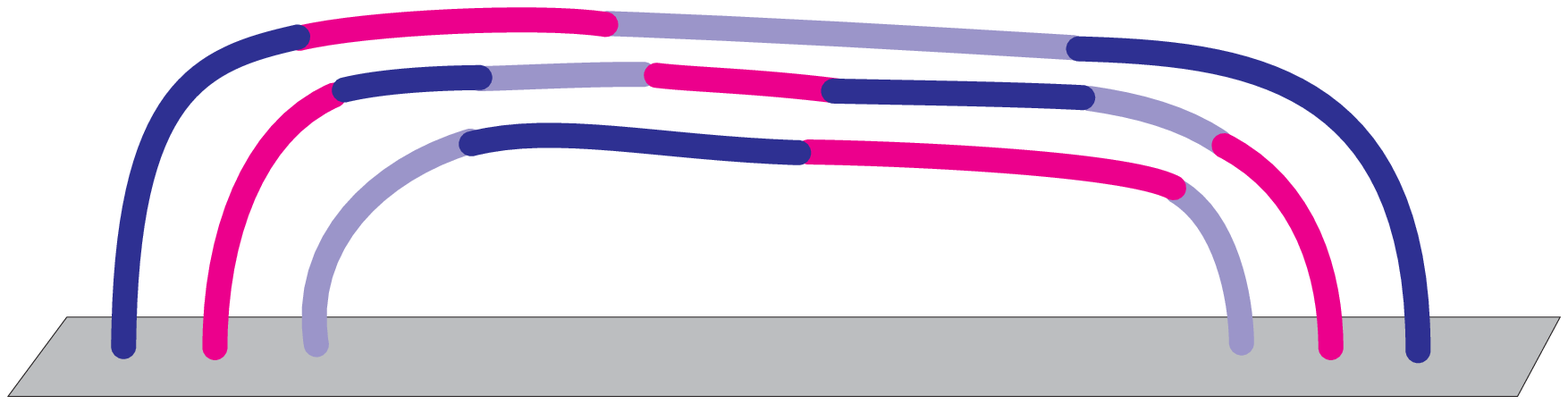,
               width=6.0cm}}
   \end{center}
 \caption{\em For the example of Fig.~\ref{fig2} a series of six
   reconnection processes leads to an unlinked field.}

 \label{fig5}
 \end{figure}

To release the above defined minimum free energy of a magnetic field
it is necessary to change its topology. The most important process for
astrophysical plasmas which allows for a change of the magnetic
topology is magnetic reconnection. Taylor conjectured (\cite{Taylor}) 
that the total helicity should be approximately conserved during a
relaxation process which involves reconnection. This conjecture
proved to be true in that the total helicity is decreasing 
on a longer time scale than the magnetic energy (\cite{Ber3}).
 Moreover, there exists a special form of reconnection events 
which exactly conserve the helicity. This is the case if 
exactly oppositely directed magnetic fields reconnect such that the
magnetic field is locally two dimensional and the electric field
is perpendicular to the magnetic field. Then ${\bf E} \!  \cdot
\!  {\bf B}=0 $ and the source term on the right hand side of the balance equation  for the helicity density,
\begin{equation} 
        \frac{\partial {\bf A}\!  \cdot \!  {\bf B}}{\partial t} 
        +  \nabla \cdot \left( \phi {\bf B} + {\bf E} \times {\bf A} \right) 
        = - 2 \ {\bf E} \!  \cdot \!  {\bf B} \label{helcons2}
\end{equation} 
vanishes. Integrated over a comoving volume with a surface 
everywhere tangential to ${\bf B}$ and on which the evolution is ideal 
${\bf E} = - {\bf v} \times {\bf B}$, Eq.~\ref{helcons2} yields
\begin{equation} 
        \frac{d}{dt} \int_V {\bf A}\!  \cdot \!  {\bf B} \ d^3r  
        = - 2 \int_V \ {\bf E} \!  \cdot \!  {\bf B} \ d^3r. \label{totalhel}
\end{equation}
Thus a reconnection process which satisfies 
${\bf E} \!  \cdot \! {\bf B}=0$ exactly conserves the total
helicity. But also for the case 
$ {\bf E} \!  \cdot \!{\bf B} \neq 0 $ the total helicity is usually a
well conserved quantity since reconnection processes in a plasma are strongly
localized and thus only a small fraction of the volume
contributes to the integral on the right hand side of
Eq.~\ref{totalhel} (\cite{Horn2}). This means that the lowest energy state accessible 
under reconnection is not the vacuum field, but a field which has the
same value of the total helicity as the initial field. This reduces 
the minimum free energy  for this kind of relaxation and consequently we have to ask whether
higher forms of linkage or knottedness will lead to additional
constraints and thus to a further reduction of the minimum free
energy. For the case of reconnection with ${\bf E} \!  \cdot \! {\bf
  B}=0$ one can easily prove that this is not the case. The argument
is that for this case magnetic reconnection is a two dimensional process
and can be represented by a simple cut and paste  of magnetic 
flux (see \cite{Freed4}). Using this picture of reconnection 
 one can transform any magnetic field consisting of a set of isolated 
linked or knotted flux tubes into a set of unlinked and unknotted flux tubes 
which have at most an internal twist corresponding to the
non-vanishing total helicity (see Fig.~\ref{fig5}). This confirms the 
original conjecture of Taylor that the total helicity is the only
conserved quantity. 

Since the higher order linkage or knottedness is not conserved for
the almost ideal magnetic reconnection with ${\bf E} \!  \cdot \! {\bf
  B}=0$, we do not expect it to be conserved for ${\bf E} \!  \cdot \! {\bf
  B} \neq 0$. However, from the above example  of (preliminary) 
topological invariants it is obvious that those quantities are
integrals of high order of the
magnetic field. They are global properties of the magnetic field and with
each integration, that is with increasing order of the invariants,
more and more information of a particular local geometry of the field is lost. On the other hand the magnetic reconnection process is a
local process driven by the local geometry of the magnetic field. Thus 
it is questionable whether the relaxation process is effective in
destroying  higher forms of knottedness and linkage. With other words,
to resolve a complicated link or knot  requires a sequence of reconnection
processes, and the order and locations of these reconnection processes 
is not arbitrary (see Fig~\ref{fig5} for an example). Therefore the relaxation might not lead at the right place in the
volume to a locally strongly sheared magnetic field, which would be
necessary to trigger a reconnection process and resolve the linkage. Moreover, the number of
reconnection processes needed to resolve a complicated topology increases with complexity. Thus these invariants, although they may
fluctuate under reconnection events, might not decay quickly in generic 
situations. Note that this does not apply to situations as in most
 technical plasmas,  where the relaxation process or the non-ideal
 region, respectively, occupies the whole volume under consideration. Only if length scales
and the magnetic Reynolds numbers are as high as in astrophysical
plasmas magnetic reconnection can be considered as a local process.
Also, one has to be cautious in applying these considerations to open
field configurations, because in this case all forms of linkage and
knottedness have to be defined with respect to a reference field
(e.g.~the vacuum field).

\section{SUMMARY}
A complex magnetic field topology can store large amounts of magnetic
energy and its minimum magnetic energy under ideal relaxations
increases with increasing complexity. Hence such field structures can
serve as an efficient energy reservoir. However, the complexity also
imposes restrictions on the release of energy from this
reservoir. This is because magnetic reconnection is the only process
capable to release this energy on a short time scale. But with respect 
to the first topological invariant, magnetic helicity,  reconnection 
turns out to be 
inefficient to relax certain configurations to their lowest energy state 
(the vacuum field), since it approximately conserves this quantity. 
This is not true for higher order topological invariants, as can be
shown easily. However, a complete relaxation in these cases requires a 
special sequence of reconnection processes, which is unlikely to
occur in real plasmas. Thus the higher forms of linkage do not
provide a strict constraint for the final state of relaxation by
reconnection, as does magnetic helicity, but they will significantly 
delay the process.

\section*{ACKNOWLEDGMENTS}
This work was  supported by the {\sl Volkswagen Foundation}.

\begin{samepage}
 
\end{samepage}

\end{document}